\documentclass[conference,a4paper]{IEEEtran}
\IEEEoverridecommandlockouts
\usepackage{amsmath}
\usepackage{amssymb}
\usepackage{amsfonts}
\usepackage{epsfig}
\usepackage{calc}
\usepackage{color}
\usepackage[all]{xy}
\usepackage{bm}
\usepackage{algorithmicx}
\usepackage{algpseudocode}
\usepackage{algorithm}
\usepackage{enumerate}
\usepackage[noadjust]{cite}
\usepackage{caption}

\newtheorem{defn}{Definition}
\newtheorem{thm}{Theorem}[section]
\newtheorem{cor}[thm]{Corollary}
\newtheorem{prop}{Proposition}

\newtheorem{lem}[thm]{Lemma}
\newtheorem{conj}[thm]{Conjecture}
\newtheorem{constr}[thm]{Construction}
\newtheorem{note}{Remark}

\newcommand{\bit}{\begin{itemize}}
\newcommand{\eit}{\end{itemize}}
\newcommand{\bcor}{\begin{cor}}
\newcommand{\ecor}{\end{cor}}
\newcommand{\beq}{\begin{equation}}
\newcommand{\eeq}{\end{equation}}
\newcommand{\beqn}{\begin{equation*}}
\newcommand{\eeqn}{\end{equation*}}
\newcommand{\bea}{\begin{eqnarray}}
\newcommand{\eea}{\end{eqnarray}}
\newcommand{\bean}{\begin{eqnarray*}}
\newcommand{\eean}{\end{eqnarray*}}
\newcommand{\ben}{\begin{enumerate}}
\newcommand{\een}{\end{enumerate}}
\newcommand{\bdefn}{\begin{defn}}
\newcommand{\edefn}{\end{defn}}
\newcommand{\bnote}{\begin{note}}
\newcommand{\enote}{\end{note}}
\newcommand{\bprop}{\begin{prop}}
\newcommand{\eprop}{\end{prop}}
\newcommand{\blem}{\begin{lem}}
\newcommand{\elem}{\end{lem}}
\newcommand{\bthm}{\begin{thm}}
\newcommand{\ethm}{\end{thm}}
\newcommand{\bconj}{\begin{conj}}
\newcommand{\econj}{\end{conj}}
\newcommand{\bconstr}{\begin{constr}}
\newcommand{\econstr}{\end{constr}}
\newcommand{\bpf}{\begin{proof}}
\newcommand{\epf}{\end{proof}}

\begin{document}
\sloppy
\title{A High-Rate MSR Code With Polynomial Sub-Packetization Level}

\author{
  \IEEEauthorblockN{Birenjith Sasidharan, Gaurav Kumar Agarwal, and P. Vijay Kumar}
  \IEEEauthorblockA{Department of Electrical Communication Engineering, Indian Institute of Science, Bangalore.\\
    Email: {biren, agarwal, vijay}@ece.iisc.ernet.in} \thanks{This research is supported in part by the National Science Foundation under Grant No. 1422955 and in part by the joint UGC-ISF Research Grant No. 1676/14. Birenjith Sasidharan would like to acknowledge the support
of TCS Research Scholar Programme Fellowship.}
}
\maketitle

\begin{abstract} We present a high-rate $(n,k,d=n-1)$-MSR code with a sub-packetization level that is polynomial in the dimension $k$ of the code. While polynomial sub-packetization level was achieved earlier for vector MDS codes that repair systematic nodes optimally, no such MSR code construction is known. In the low-rate regime (i. e., rates less than one-half), MSR code constructions with a linear sub-packetization level are available. But in the high-rate regime (i. e., rates greater than one-half), the known MSR code constructions required a sub-packetization level that is exponential in $k$. In the present paper, we construct an MSR code for $d=n-1$ with a fixed rate $R=\frac{t-1}{t}, \ t \geq 2,$ achieveing a sub-packetization level $\alpha = O(k^t)$. The code allows help-by-transfer repair, i. e., no computations are needed at the helper nodes during repair of a failed node.
\end{abstract}


\section{Introduction\label{sec:intro}}

In a distributed storage system, the data file comprising of $B$ data symbols drawn from a finite field $\mathbb{F}_q$, is encoded using  an error-correcting code of block length $n$ and the code symbols are stored in $n$ nodes of the storage network.  A naive strategy aimed at achieving resilience against node failures is to store multiple replicas of the same data. Given the massive amount of data being stored, sophisticated codes such as Reed-Solomon (RS) codes with low storage overhead are being employed in practice. However, the amount of data download required to repair a single node-failure is quite large for the RS codes. The framework of regenerating codes was introduced in~\cite{DimGodWuWaiRam} to address this problem. In an $(n,k,d)$-regenerating code, a file comprised of $B$ symbols from a finite field $\mathbb{F}_q$ is encoded into a set of $n\alpha$ coded symbols and then stored across $n$ nodes in the network with each node storing $\alpha$ coded symbols.  The parameter $\alpha$ is termed as the {\em sub-packetization} level of the code. A data collector can download the data by connecting to any $k$ nodes.  In the event of node failure, node repair is accomplished by having the replacement node connect to any $d$ nodes and download $\beta \leq \alpha$ symbols from each node with $\alpha \leq d \beta < B$. The quantity $d\beta$ is termed the {\em repair bandwidth}. Here one makes a distinction between functional and exact repair.  By {\em functional repair} (FR), it is meant that a failed node will be replaced by a new node such that the resulting network continues to satisfy the data collection and node-repair properties defining a regenerating code.   An alternative to function repair is {\em exact repair} (ER) under which one demands that the replacement node store precisely the same content as the failed node. 

A cut-set bound based on network-coding concepts, tells us that given a code parameter set $(n,k,d)$, the maximum possible size of a data file under FR, is upper bounded~\cite{DimGodWuWaiRam} by 
\vspace{-0.2in}
\bea \label{eq:cut_set_bd}
B & \leq & \sum_{\ell =1}^{k} \min\{\alpha,(d-\ell +1)\beta\} .
\eea
The above bound is tight since the existence of codes achieving this bound has been established using network-coding arguments related to multicasting.  For fixed values of $(n,k,d,B)$, the bound in \eqref{eq:cut_set_bd} characterizes a tradeoff between $\alpha$ and $\beta$, referred to as the Storage-Repair Bandwidth tradeoff. The two extremal points in the tradeoff are respectively, the minimum-storage regenerating (MSR) and minimum bandwidth regenerating (MBR) points which correspond to the points at which the storage and repair bandwidth are respectively minimized. At MBR point, we have
\beq
\alpha \ = \ d\beta, \ B \ = \ k\alpha - {k \choose 2}\beta,
\eeq
and at MSR point, we have
\beq
\alpha \ = \ (d-k+1)\beta, \ B \ = \ k\alpha.
\eeq
It is proved that MSR and MBR points are achievable by ER codes as well. The focus of the current paper is on ER MSR codes and for convenience we simply refer to them as MSR codes. 

\subsection{MSR Codes}

The MSR codes can be considered as codes over a vector alphabet $\mathbb{F}_{q^{\alpha}}$ with dimension $k$. Since they tolerate any $(n-k)$ node-erasures, and they have a file size of $B=k\alpha$, MSR codes are Maximum-Distance-Separable (MDS) codes over the vector alphabet $\mathbb{F}_{q^{\alpha}}$. The combination of these two properties is therefore called the {\em MDS property} of MSR codes. On the other hand, MSR codes in addition to being vector MDS codes can repair a failed node with the least possible repair bandwidth. 

The construction of MSR codes is a well-studied problem in literature. In~\cite{RasShaKum_pm}, a framework to construct MSR codes is provided for $d \geq 2k-2$. In~\cite{PapDimCad}, high-rate MSR codes with parameters $(n, k=n-2, d=n-1$) are constructed using Hadamard designs. In~\cite{TamWanBru}, high-rate MSR codes, known as zigzag codes, are constructed for $d = n-1$; here efficient node-repair is guaranteed only in the case of systematic nodes. This was subsequently extended to include the repair of parity nodes as well in~\cite{WanTamBru_allerton}. A construction for MSR codes with $d = n -1 \geq 2k - 1$ using techniques of interference alignment is presented in~\cite{SuhRam} and \cite{ShaRasKumRam_ia}. In \cite{CadJafMalRamSuh}, authors showed the existence of MSR codes for any value of $(n,k,d)$.

\subsection{Our Approach On Sub-packetization and Contributions}


A parameter of interest for MSR codes is the amount of sub-packetization ($\alpha$) required for a given value of $(n,k,d)$. The MSR constructions known as zigzag codes that allow arbitrarily high rates required a sub-packetization level that is exponential in $k$. Later in~\cite{WanTamBru_long}, a vector MDS codes that repair systematic nodes was constructed achieving $\alpha = r^{\frac{k}{r+1}}$ where $r:=n-k$. Recently in \cite{AgaSasKum}, another vector MDS code that repairs systematic nodes optimally was proposed satisfying an additional property known as {\em access-optimality}. The construction required $\alpha = r^{\frac{k}{r}}$. In ~\cite{GopTamCal}, authors derived a lower bound on the sub-packetization in terms of $k$, and $r$ as given below:
\bean
2\log_2 \alpha (\log_{\left(\frac{r}{r-1}\right)} \alpha +1) + 1 & \geq & k.
\eean
Earlier in~\cite{CadHuaLiSan}, authors constructed a vector MDS code with rate $R=\frac{2}{3}$, requiring an $\alpha$ that is polynomial in $k$. They could also achieve polynomial $\alpha$ for any fixed rate in the regime $\frac{2}{3} \leq R \leq 1$. However, these codes were also limited by the fact that optimal repair was feasible for systematic codes alone. Quite similar to the approach in \cite{CadHuaLiSan}, we also restrict our focus to the family of MSR codes with a fixed rate $R = \frac{t-1}{t}, t \geq 2$. It is worthwhile to remark at this point that the family of Product-Matrix MSR codes~\cite{RasShaKum_pm} with rate restricted by $R \leq \frac{1}{2}$ required only a linear sub-packetization level. In the present paper, we construct a $(n,k,d=n-1)$-MSR code with a fixed rate $R = \frac{t-1}{t}$ where $t \geq 2$ is an integer parameter. The code will have $\alpha = \left(\frac{k}{t}\right)^t$. To the best of our knowledge, these are the first MSR constructions that achieve a sub-packetization level that is polynomial in $k$. These codes are help-by-transfer codes, by which we mean that the helper nodes need not do any computation during the repair of a failed node. 

\section{MSR Code Construction For Rate$=\frac{t-1}{t}$ \label{sec:msr-code}}

In this section, we provide the construction for MSR codes with a rate, $R=\frac{t-1}{t}$ for some positive integer $t$. The construction is described for a particular example of $t=3$, and subsequently generalized. 

\subsection{Code Construction for $R=\frac{2}{3}$\label{sec:2by3}}

We have an auxiliary parameter $q=p^m$ for some prime $p$, and $m$ a positive integer\footnote{The auxiliary parameter takes values from a finite-field, though it is sufficient to work with a finite-ring. This does not cause any lack of generality in the principles used for the construction.}. Then the code has parameters
\bean
n \ = \ 3q, \ k \ = \ 2q, \ d \ = \ (n-1), \ \alpha \ = \ q^3 .
\eean
A codeword of an MSR code can be treated as an array of size $(\alpha \times n)$. We first introduce an indexing for the rows and columns (nodes and columns are often used interchangeably) of the codeword array. Let $\mathbb{F}_q = \{0,1,\ldots, q-1\}$ denote a finite field of size $q$, and a $2$-tuple $(i,\theta), i \in \{1,2,3\}, \ \theta \in \mathbb{F}_q $ is used to index the columns. The rows are indexed by elements $(x,y,z)$ from $\mathbb{F}_q^3$ where $x,y,z \in \mathbb{F}_q$. Thus $C(x,y,z; (i,\theta))$ represents one code symbol from the codeword array at the intersection of the row $(x,y,z)$ and the node $(i,\theta)$. In order to describe the code, we first introduce the following notation 
\bea
\label{not:oplus}
a_1 \oplus a_2 \oplus \dots \oplus a_n := \sum\limits_{i=1}^{n} {c_i.a_i}, \ \  c_i \neq 0, \ \forall i \in [n]
\eea
to denote a linear combination involving each of the scalars in $\{a_1, a_2, \ldots, a_n\}$ with non-zero coefficients. The notation is oblivious to the particular choice of non-zero coefficients in the linear combination. The code is described by $q^4$ parity-check constraints. Throughout this paper, the symbol $\sum$ is used with a different meaning. The terms within a $\sum$ are not connected by the binary operator $+$, but by the $\oplus$ operator as defined in \eqref{not:oplus}.  For every $(x,y,z) \in \mathbb{F}_q^3$,
\beqn
\sum_{\theta \in \mathbb{F}_q} C(x,y,z; (1,\theta)) \ \oplus \ \sum_{\theta \in \mathbb{F}_q} C(x,y,z; (2,\theta)) \ \oplus \eeqn \beq
 \label{eq:pc1} \sum_{\theta \in \mathbb{F}_q} C(x,y,z; (3,\theta))  \ = \ 0,
\eeq
\beqn
C(x-\Delta,y,z; (1,x)) \ \oplus \ C(x,y-\Delta,z; (2,y)) \ \oplus  \eeqn
\beqn C(x,y,z-\Delta; (3,z)) \  \oplus \ \sum_{\theta \in \mathbb{F}_q} C(x,y,z; (1,\theta)) \  \oplus \eeqn \beq
\sum_{\theta \in \mathbb{F}_q} C(x,y,z; (2,\theta)) \ \oplus \sum_{\theta \in \mathbb{F}_q} C(x,y,z; (3,\theta)) \ =\  0, \ \Delta \in \mathbb{F}_q^* \label{eq:pc2} .
\eeq
The parity-check constraint in \eqref{eq:pc1} is referred to as the {\em row-parity}, and the that in \eqref{eq:pc2} is referred to as the {\em $\Delta$-parity}. It can be observed that the first three terms in the $\Delta$-parity equations are entries that do not belong to the $(x,y,z)$-row. These entries are referred to as the {\em shifted entries}. What remains is the identification of coefficients in these parity-check constraints so that the MDS property holds. Instead of constructing these coefficients explicitly, we will show in Sec.~\ref{sec:mds} that such coefficients indeed exist in a sufficiently large field. Therefore, the description of the code is complete with \eqref{eq:pc1}, \eqref{eq:pc2}.

In the zigzag code~\cite{TamWanBru}, parity symbols are categorized into two types, namely row-parities and zigzag parities. The row parities are made up of message symbols from the same row of the codeword array. But the zigzag parities are made up of message symbols belonging to various rows such that one message symbol is picked per column. In our construction also, every parity-check constraint corresponding to $\Delta \neq 0$ involves shifted entries that do not belong to the row under consideration. In this manner, our construction is of a similar flavor as that in \cite{TamWanBru}. But the major difference of our construction from the zigzag construction lies in the symmetry of the parity-check constraints. It also differs in the fact that two symbols of the same column can be involved in the same parity-check constraint in the case of $\Delta \neq 0$. Such an approach was earlier adopted in~\cite{AgaSasKum}.

\subsubsection{Optimal Repair of a Failed Node \label{sec:repair}}

Without loss of generality, assume that the node $(1,\theta_0)$ failed. We download symbols belonging the rows $\Gamma = \{(\theta_0,y,z) \mid y,z \in \mathbb{F}_q \}$. Clearly $|\Gamma| = q^2$.  Thus we have $\{C(\theta_0,y,z;(i,\theta)) \mid i=1,2,3, \theta \neq \theta_0, y,z \in \mathbb{F}_q \}$. The rows are selected such that $x = \theta_0$, because the first coordinate of the index of the node is $1$. If the first coordinate had been $2$ or $3$, we would have fixed $y=\theta_0$ or $z=\theta_0$ respectively. All the code symbols 
\bean
C(\theta_0,y,z;(1,\theta_0)), \ \ y,z \in \mathbb{F}_q
\eean
are repaired using the row-parities. Hence we have all the symbols belonging to rows in $\Gamma$ from all the $n$ nodes. Next, let us write the equation for $\Delta$-parity, $\Delta \in \mathbb{F}_q^*$ corresponding to an arbitrary row $(\theta_0,y,z) \in H$.
\bea
\nonumber C(\theta_0-\Delta,y,z; (1,\theta_0)) \ \oplus \ C(\theta_0,y-\Delta,z; (2,y)) \ \oplus \\
\nonumber C(\theta_0,y,z-\Delta; (3,z)) \oplus \sum_{\theta \in \mathbb{F}_q} C(\theta_0,y,z; (1,\theta)) \ \oplus \\
\label{eq:repair1}  \sum_{\theta \in \mathbb{F}_q} C(\theta_0,y,z; (2,\theta)) \ \oplus \ \sum_{\theta \in \mathbb{F}_q} C(\theta_0,y,z; (3,\theta)) \ = \  0.
\eea

Except the term $C(\theta_0-\Delta,y,z; (1,\theta_0))$, all other symbols involved in \eqref{eq:repair1} are known to us. Thus  $C(\theta_0-\Delta,y,z; (1,\theta_0))$ can be repaired for all choices of $y,z$. By making use of all the $\Delta$-parities, we can thus repair all the remaining symbols in the node $(1,\theta_0)$. The total number of symbols downloaded per node is
\bean
\beta & = & q^2 \ = \ \frac{q^3}{q} \ = \ \frac{\alpha}{d-k+1},
\eean
and thus the repair is bandwidth-optimal.

\subsubsection{The MDS Property\label{sec:mds}}

In this section, we will show that we can find an assignment of coefficients to the row-parities and $\Delta$-parities such that the code satisfies the MDS property. We start with stating a useful fact. 
\blem \label{lem:fact}Let $H$ be a $((n-k)\times n)$-parity-check matrix of a linear code ${\cal C}$. If $S \subset [n], |S|=(n-k)$ is such that $\textsl{rank}(H\mid_S) = (n-k)$, then it is possible to decode every codeword of ${\cal C}$ accessing symbols belonging to locations $S^c = [n] \setminus S$.
\elem

Based on the parity-constraints in \eqref{eq:pc1}, \eqref{eq:pc2}, we will determine the structure of the parity-check matrix $H$. First,  we vectorize the codeword array node-by-node so that the first $\alpha=q^3$ columns of $H$ represent the first node, the second $q^3$ columns represent the second node and so on. The group of $q^3$ columns associated with a node is referred to as a {\em thick column}. The parity-check matrix thus obtained will be of size $(q^4 \times 3q^4)$ with $n$ thick columns each containing $\alpha$ {\em thin columns}. In order to describe the support and thereby the structure of $H$, we will for a moment assume that all the coefficients are set to $1$. This matrix is denoted by $H_{s}$, and is given by
\bean
H_{s} & = & J + E, 
\eean
where the matrices $J$ and $E$ are given in \eqref{eq:J} and \eqref{eq:E}. The equation~\eqref{eq:J} also illustrates the fact that the rows of $J$ can be decomposed into blocks of size $q^3$, each corresponding to parity-check constraints with a fixed $\Delta$. The first set of $q^3$ parity-check constraints correspond to row-parities possibly associated with $\Delta=0$.

\begin{figure*}
\bea
\label{eq:J} J & = &  \begin{array}{c} \Delta = 0 \\ \hline  \Delta = 1 \\ \hline \vdots \\  \\ \hline \Delta = q-1  \end{array}   \left[\begin{array}{ccc|ccc|ccc} I_{q^3} & \cdots & I_{q^3}    &   I_{q^3} & \cdots & I_{q^3}   &   I_{q^3} & \cdots & I_{q^3} \\
\hline
I_{q^3} & \cdots & I_{q^3}    &   I_{q^3} & \cdots & I_{q^3}  &    I_{q^3} & \cdots & I_{q^3} \\
\hline
\vdots  & \vdots &     &    &  &   &   & \vdots &   \\
  &     &    &  &   &   &        &  \\
\hline
I_{q^3} & \cdots & I_{q^3}    &   I_{q^3} & \cdots & I_{q^3}  &   I_{q^3} & \cdots & I_{q^3} 
\end{array} \right]
\eea
\bea
\label{eq:E} E & = & \left[\begin{array}{ccc|ccc|ccc} 0_{q^3} & \cdots & 0_{q^3}    &   0_{q^3} & \cdots & 0_{q^3}    &   0_{q^3} & \cdots & 0_{q^3} \\
\hline
E^{1}_{1,1} & \cdots & E^{1}_{1,q}    &   E^{2}_{1,1} & \cdots & E^{2}_{1,q}   &   E^{3}_{1,1} & \cdots & E^{3}_{1,q} \\
\hline
 & \vdots &     &    &  &   &  & \vdots &   \\
 &        &     &    &  &   &  &        &  \\
\hline
E^{1}_{q-1,1} & \cdots & E^{1}_{q-1,q}    &   E^{2}_{q-1,1} & \cdots & E^{2}_{q-1,q}    &  E^{3}_{q-1,1} & \cdots & E^{3}_{q-1,q} \\
\end{array} \right].
\eea
\hrulefill
\end{figure*}

In \eqref{eq:J}, \eqref{eq:E}, $I_{q^3}, 0_{q^3}$ respectively represent identity and all-zero matrix of size $q^3 \times q^3$. The matrices $\{ E^{i}_{\delta,\theta} \mid i \in \{1,2,3\}, \delta \in \mathbb{F}^*_q, \theta \in \mathbb{F}_q \}$ are made up of $1$s and zeros, and represent the shifted entries of the corresponding $\Delta$-parities. The matrices $J, E$ and hence $H$ are block matrices of size $(q \times 3q)$ where each block is a square matrix of size $q^3$. We will show that the MDS property can be ensured by assigning suitable coefficients to locations identified by the support of $H_s$. Our method is quite similar to the method used in~\cite{AgaSasKum}. By~\ref{lem:fact}, it is sufficient that $H$ restricted to any $(n-k)=q$ thick columns has a rank equal to $q\alpha=q^4$. Let us assign an indeterminate $c$ to all the locations determined by the support of $E$. Now consider the square submatrix $H_D$ obtained by restricting $H$ to $D \subset [n], |D| = (n-k)$ thick columns. If we assume that all the coefficients of $J$ are fixed, the determinant of $H_D$ will be a polynomial in the indeterminate $c$. Let us denote this polynomial by $p_D(c)$. In the following lemma, we prove that $p_D(c)$ can be made a non-zero polynomial for every choice of $D \subset [n], |D| = n-k$.
\blem \label{lem:J} There exists an assignment of coefficients to $J$ such that $p_D(c)$ is a non-zero polynomial for every choice of $D \subset [n], |D| = n-k$. 
\elem
\bpf Consider a $[3q,2q]$-RS code and its parity-check matrix $H_{\text{mds}}$ of size $(q \times 3q)$. Clearly a $(q\times q)$-matrix obtained by restricting $H_{\text{mds}}$ to any $q$ columns has full rank. Let $A \otimes B$ denote the Kronecker product of matrices $A$ and $B$. If we set $J$ to $J_0$,
\bea \label{eq:j0}
J_0 & = & H_{\text{mds}} \otimes I_{q^3},
\eea
then we must have $p_D(0)$ evaluating to a non-zero value for every choice of $D \subset [n], |D| = n-k$. Hence $p_D(c)$ must be a non-zero polynomial for every valid choice of $D$.
\epf
Henceforth, we assume that the coefficients of the polynomials $p_D(c)$ are fixed by the coefficients of $J$ as determined by Lemma~\ref{lem:J}. By the structure of $E$, it is clear that 
\beq
\textsl{deg}(p_D(c)) \leq q^4 - q^3 .
\eeq
Next, consider the polynomial
\bea
p(c) & = & \prod_{D \subset [n], |D|=k} p_D(c).
\eea
Clearly $p(c)$ is not identically zero, and its degree is upper bounded by ${n \choose k}q^3(q-1)$. Hence it is sufficient that we find an non-zero assignment $c_0 \neq 0$ for $c$ such that $p(c_0) \neq 0$. By Combinatorial Nullstellansatz~\cite{Alo}, this is possible if we choose the field size greater than ${n \choose k}q^3(q-1) + 1$. Thus we have proved the following theorem.
\bthm \label{thm:mds}There exists an assignment for the coefficients in the parity-check constraints in~\eqref{eq:pc1},~\eqref{eq:pc2} such that the code described in~\ref{sec:2by3} is an MSR code.
\ethm
Using the constant $c_0$ guaranteed in the proof of Thm.~\ref{thm:mds}, and $J_0$ in \eqref{eq:j0}, the parity-check matrix $H$ of the MSR code takes the form
\bea \label{eq:tpcmatrix} H & = & J_0 + c_0 E.
\eea


\subsection{Code Construction for $R=\frac{t-1}{t}, t \geq 2$\label{sec:byt}}

The principle of the construction is elucidated in the last section completely, and the generalization to the case of rate $R = \frac{t-1}{t}, t \geq 2$ is straightforward.  For an auxiliary parameter $q=p^m$ for some prime $p$, and $m$ a positive integer, the code construction has parameters
\bean
n \ = \ tq, \ k \ = \ (t-1)q, \ d \ = \ (n-1), \ \alpha \ = \ q^t .
\eean
A $2$-tuple $(i,\theta), i \in \{1,2,\ldots, t\}, \ \theta \in \mathbb{F}_q $ is used to index the columns. The rows are indexed by elements $(x_1,x_2,\ldots, x_t)$ from $\mathbb{F}_q^t$ where $x_j \in \mathbb{F}_q$. Thus $C(x_1,x_2,\ldots, x_t; (i,\theta))$ represents one code symbol from the codeword array at the intersection of the row $(x_1,x_2,\ldots, x_t)$ and the node $(i,\theta)$. The code is described by $q^{t+1}$ parity-check constraints. For every $\underline{x} = (x_1,x_2,\ldots, x_t) \in \mathbb{F}_q^t$,
\bea 
\nonumber \sum_{\theta \in \mathbb{F}_q} C(\underline{x}; (1,\theta)) \oplus \sum_{\theta \in \mathbb{F}_q} C(\underline{x}; (2,\theta)) \oplus \\
\label{eq:tpc1} \cdots \ \oplus \sum_{\theta \in \mathbb{F}_q} C(\underline{x}; (t,\theta))  \ = \ 0,
\eea
\beqn
C(x_1-\Delta,x_2,\ldots, x_t; (1,x_1)) \ \oplus \ C(x_1,x_2-\Delta,\ldots, x_t; (2,x_2)) \ \oplus \\
\eeqn
\beqn
\cdots  \ \oplus \ C(x_1,x_2,\ldots, x_t-\Delta; (t,x_t)) \ \oplus \ \sum_{\theta \in \mathbb{F}_q} C(\underline{x}; (1,\theta)) \ \oplus
\eeqn \beq \sum_{\theta \in \mathbb{F}_q} C(\underline{x}; (2,\theta)) \ \oplus \ \cdots \oplus \sum_{\theta \in \mathbb{F}_q}  C(\underline{x}; (t,\theta)) \ = \  0, \ \Delta \in \mathbb{F}_q^* . \label{eq:tpc2} 
\eeq
The parity-check constraint in \eqref{eq:tpc1} is referred to as the {\em row-parity}, and the that in \eqref{eq:tpc2} is referred to as the {\em $\Delta$-parity}. As in the special case of $t=3$ described in Sec.~\ref{sec:2by3}, the first $t$ terms in the $\Delta$-parity equations are entries that do not belong to the $(x_1,x_2,\ldots,x_t)$-row. These entries are referred to as the {\em shifted entries}. Existence of coefficients for parity-check equations that ensure MDS property follows in the same line as that in Sec.~\ref{sec:mds}. What remains is to present a repair strategy that is bandwidth-optimal.

\subsubsection{Optimal Repair of a Failed Node \label{sec:repair}}

Assume that the node $(i_0,\theta_0)$ failed. We download symbols belonging the rows $\Gamma = \{\underline{x} \mid x_j \in \mathbb{F}_q, \forall j \neq i_0, \ x_{i_0} = \theta_0 \}$. Clearly $|\Gamma| = q^{t-1}$. Thus we have $\{C(\underline{x};(i,\theta)) \mid (i,\theta) \neq (i_0,\theta_0), \underline{x} \in \Gamma \}$. All the code symbols 
\bean
C(\underline{x};(i_0,\theta_0)), \ \ x \in \Gamma
\eean
are repaired using the row-parities. Then we have all the symbols belonging to rows in $\Gamma$ from all the $n$ nodes. Next, let us write the equation for $\Delta$-parity, $\Delta \in \mathbb{F}_q^*$ corresponding to an arbitrary row $\underline{x} \in \Gamma$.
\beqn
C(x_1-\Delta,x_2,\ldots,x_t; (1,x_1)) \ \oplus \ \cdots \ \oplus \eeqn \beqn
C(x_1,\ldots,x_{i_0}-\Delta,\ldots,x_t; (i_0,\theta_0)) \ \oplus \ \cdots \ \oplus \eeqn 
\beq 
\label{eq:repair2}
C(x_1,x_2,\ldots,x_t-\Delta; (t,x_t)) \ \oplus \ \sum_{\theta \in \mathbb{F}_q , j \in [t]} C(\underline{x}; (j,\theta))  \ = \  0.
\eeq
Except the term $C(x_1,\ldots,x_{i_0}-\Delta,\ldots,x_t; (i_0,\theta_0))$, all other symbols involved in \eqref{eq:repair2} are known to us. By varying $\Delta$, we can thus repair all the remaining symbols in the node $(i_0,\theta_0)$. The total number of symbols downloaded per node is
\bean
\beta & = & q^{t-1} \ = \ \frac{q^t}{q} \ = \ \frac{\alpha}{d-k+1},
\eean
and thus the repair is bandwidth-optimal.

\bibliographystyle{IEEEtran}
\bibliography{isit2015}
\end{document}